\documentclass[11pt,a4paper,english,showpacs,superscriptaddress,prd,aps,preprint]{revtex4-1}
\usepackage{amssymb}
\usepackage{amsmath} 
\usepackage{graphicx}
\usepackage{babel}
\usepackage{hyperref}
\usepackage{color}

\begin{document}

\title{Gravitational corrections to the scattering in a massless quantum electrodynamics}

\author{B.~Charneski}
\email{bruno@fma.if.usp.br}
\affiliation{Instituto de F\'\i sica, Universidade de S\~ao Paulo\\
Caixa Postal 66318, 05315-970, S\~ao Paulo, S\~ao Paulo, Brazil}

\author{A.~C.~Lehum}
\email{andrelehum@ect.ufrn.br}
\affiliation{Instituto de F\'\i sica, Universidade de S\~ao Paulo\\
Caixa Postal 66318, 05315-970, S\~ao Paulo, S\~ao Paulo, Brazil}
\affiliation{Escola de Ci\^encias e Tecnologia, Universidade Federal do Rio Grande do Norte\\
Caixa Postal 1524, 59072-970, Natal, Rio Grande do Norte, Brazil}

\author{A. J. da Silva}
\email{ajsilva@fma.if.usp.br}
\affiliation{Instituto de F\'\i sica, Universidade de S\~ao Paulo\\
Caixa Postal 66318, 05315-970, S\~ao Paulo, S\~ao Paulo, Brazil}

\begin{abstract}

We discuss the role of gravitational corrections to the running of the electric charge through the evaluation of scattering amplitudes of charged particles in massless scalar electrodynamics. Computing the complete divergent part of the S-matrix amplitude for two distinct scattering processes, we show that quantum gravitational corrections do not alter the running behavior of the electric charge. Our result does not exclude the possibility that the presence of a second dimensional constant in the model (a cosmological constant or the presence of massive particles) could alter this behavior, as was proposed in earlier works.
\end{abstract}



\maketitle

\section{Introduction}

The theory of gravitational field quantized for small fluctuations around a flat metric is nonrenormalizable~\cite{'tHooft:1974bx,PhysRevLett.32.245,Deser:1974cy}, i.e., an infinite number of free parameters is required to absorb all kinds of new UV divergences which are generated, as the order of perturbative calculations is increased. On the other hand, quantum effects due to gravitation at low energy, much below the Planck scale $M_P\approx 1.4 \times 10^{19}$ GeV/c$^2$ (a natural energy scale of the quantum theory of gravitation), can be calculated by incorporating its effects into the spirit of an effective field theory~\cite{Donoghue:1994dn}. From this point of view, Robinson and Wilczek~\cite{Robinson:2005fj} proposed that quantum gravity corrections make gauge theories asymptotically free (i.e., the gauge coupling constant goes to zero in the limit of very high energy scales), even though the gauge coupling does not exhibit this property in the absence of gravity [as in quantum electrodynamics (QED)]. The origin of this effect would be the arising of quadratic UV divergences, associated to the one graviton exchange graph, that could be absorbed in a gauge coupling constant redefinition. 

After Robinson and Wilczek's paper, some authors have argued that this effect should be gauge dependent~\cite{Pietrykowski:2006xy} and ambiguous~\cite{Felipe:2012vq}-- and therefore without physical meaning. Considering the QED coupled to gravity and using the Vilkovisky-DeWitt method, which is a gauge-invariant and gauge-condition-independent way of computing effective action, it was reinforced that the quadratic divergences, due to the gravitational corrections, turn the electric charge asymptotically free~\cite{Toms:2010vy,Toms:2011zza}. Later in~\cite{Nielsen:2012fm}, it was shown, for the Einstein-Maxwell system, that the Vilkovisky-DeWitt method guarantees only the gauge invariance of finite and logarithmic divergent parts of the effective action; quadratic divergences would break the Ward identities.

 It has also been argued in~\cite{Toms:2008dq} that asymptotic freedom in gauge theories could also happen in the presence of a cosmological constant, through logarithmic divergences, made possible by the presence of a second dimensionful parameter ($\Lambda$), besides $M_P$. In addition, in nongauge theories, the Yukawa and $\phi^4$ interactions were shown to share the same property of asymptotic freedom~\cite{Rodigast:2009zj}, due to gravitational corrections, in the presence of mass for the scalar and fermionic particles, an effect that vanishes when the masses are withdrawn.  

On the other hand, the physical significance of the definition of the running coupling constants, as inferred from the effective action, was questioned by Anber et. al.~\cite{Anber:2010uj} and Donoghue~\cite{Donoghue:2012zc}; as argued by them, only a scattering matrix computation should give a real physical definition for the running of the coupling constants. Through the computation of scattering amplitudes in the presence of quantum gravity, they have shown that attempts to define a running Yukawa coupling would be process dependent; i.e., what appears as asymptotic freedom behavior in one process would appear in another as an increase of the strength of the coupling constant~\cite{Anber:2010uj}. So, what appears to be asymptotic freedom is in fact a process dependent result and not a universal behavior that can be summarized in a definition of the physical coupling constant.

In view of these controversial results, in a tentative attempt to shed some light on this problem, we study in the present paper the scalar QED involving two different massless ÒpionsÓ that are coupled to gravity. We evaluate the gravitational contribution to the running of the electric charge and the $\phi^4$ self interaction coupling constant, through a direct computation of scattering amplitudes. This is done for two different processes, involving the two charged pions. We show that quantum gravitational corrections do not contribute to the running of the electric charge and the $\phi^4$ coupling. 

In summary, our conclusions agree with that of the authors of Refs.~\cite{Anber:2010uj} and \cite{Donoghue:2012zc} which, through a proper definition of the physical electric charge and the  $\phi^4$ coupling constant, indicate that their running is not altered by quantum gravitational corrections (up to the order of one gravition exchange), as originally suggested in~\cite{Robinson:2005fj}. 

The article is organized as follows: in Sec.\ref{sec1} we present the model, a massless scalar electrodynamics with two different massless scalar particles (pions) coupled to the Einstein gravity, expand it around a flat metric and compute its propagators. In Sec.\ref{sec2}, using dimensional regularization (DR) and minimal subtraction \cite{minsub}, we evaluate the divergent part of the scattering amplitudes  ($\pi_a^+ +\pi_b^+\rightarrow \pi_a^+ +\pi_b^+$) and  ($\pi_a^+ +\pi_a^-\rightarrow \pi_b^+ +\pi_b^-$). By using renormalization by minimal subtraction \cite{minsub}, we show that quantum gravitational corrections only renormalize higher order operators and have no effect on the running of the electric charge and the $\phi^4$ coupling constant. As is already known, DR automatically renormalizes quadratic UV divergences. Anyway, as discussed in~\cite{Anber:2010uj}, quadratic divergences that would appear in other regularization schemes, do not have any role in the running of the coupling constants.

\section{Scalar QED coupled to Gravity}\label{sec1}

The effective model for charged pions in electromagnetic interaction (scalar QED) coupled to the Einstein gravity is given by the following action
\begin{eqnarray}\label{eq01}
S=\int{d^4x }\sqrt{-g}&&\Big{\{}\frac{2}{\kappa^2}R-\frac{1}{4}g^{\mu\alpha}g^{\nu\beta}F_{\alpha\beta}F_{\mu\nu}+g^{\mu\nu}\left( \partial_\mu +ieA_\mu\right) \phi_j \left(\partial_\nu-ieA_\nu\right)\phi^*_j\nonumber\\
&&-\frac{\lambda}{2}(\phi^*_j\phi_j)^2 +\mathcal{L}_{HO} +\mathcal{L}_{GF}+\mathcal{L}_{CT} \Big{\}},
\end{eqnarray}

\noindent where $\kappa^2=32\pi G=32\pi/M_P^2$, with $M_P$ being the Planck mass and $G$ the Newtonian gravitational constant, 
$e$ the electric charge of the pions and $\lambda$ a self-interaction constant (already present in the scalar electrodynamics, in the absence of gravitation). $\mathcal{L}_{HO}$ is the Lagrangian of higher derivatives monomials, necessary to compensate higher order infinities induced by the nonrenormalizable gravitational interactions~\cite{Donoghue:1994dn}, $\mathcal{L}_{GF}$ is the gauge fixing plus Faddeev-Popov ghost Lagrangian (for the graviton and the photon) and $\mathcal{L}_{CT}$ is the Lagrangian of counterterms.  
We  work in the context of renormalized perturbation theory, so all coupling constants are the physical ones.  

The reason for considering two flavors of pions ($\phi_a$ and $\phi_b$, i.e., $j=a,b$) is to simplify the calculations. By studying the scattering of nonidentical particles, the scattering amplitudes get reduced to only one reaction channel, which reduces the (high) number of graphs involved.

We will consider small fluctuations around the flat metric, i.e.,  
\begin{eqnarray}\label{eq02}
&&g_{\mu\nu}=\eta_{\mu\nu}+\kappa h_{\mu\nu},\\
&&g^{\mu\nu}=\eta^{\mu\nu}-\kappa h^{\mu\nu}+\kappa^2 h^{\mu\alpha}{h_{\alpha}}^{\nu}+\mathcal{O}(\kappa^3),\\
&&\sqrt{-g}=1+\frac{1}{2}\kappa h-\frac{1}{4}\kappa^2h_{\alpha\beta}P^{\alpha\beta\mu\nu}h_{\mu\nu}+\mathcal{O}(\kappa^3),
\end{eqnarray}
\noindent where $\eta_{\mu\nu}=(+,-,-,-)$, $P^{\alpha\beta\mu\nu}=\dfrac{1}{2}(\eta^{\alpha\mu}\eta^{\beta\nu}+\eta^{\alpha\nu}\eta^{\beta\mu}-\eta^{\alpha\beta}\eta^{\mu\nu})$ and $h=\eta^{\mu\nu}h_{\mu\nu}$. For more details in obtaining the expanded Lagrangian, see, for instance~\cite{Choi:1994ax} .  

Employing the harmonic gauge-fixing function, $G_\mu=\partial^\nu h_{\mu\nu}-\dfrac{1}{2}\partial_\mu h$, by adding to the Lagrangian the term $\xi_h G^2/2$ plus the corresponding Faddeev-Popov ghost term, the graviton propagator can be cast as
\begin{eqnarray}\label{eq04}
\langle T~h^{\alpha\beta}(p) h^{\mu\nu}(-p)\rangle&=&D^{\alpha\beta\mu\nu}(p)=\frac{i}{p^2}\left[P^{\alpha\beta\mu\nu}+(\xi_h-1)\frac{Q^{\alpha\beta\mu\nu}(p)}{p^2}
\right],
\end{eqnarray}
\noindent where $Q^{\alpha\beta\mu\nu}(p)=(\eta^{\alpha\mu}p^{\beta}p^\nu+\eta^{\alpha\nu}p^{\beta}p^\mu+\eta^{\beta\mu}p^{\alpha}p^\nu+\eta^{\beta\nu}p^{\alpha}p^\mu)$.

The propagators for the other fields are also obtained by the usual Faddeev-Popov method, resulting in
\begin{eqnarray}\label{eq05}
\langle T~A^\mu(p) A^\nu(-p)\rangle&=&\Delta^{\mu\nu}(p)=-\frac{i}{p^2}\left[ \eta^{\mu\nu}+(1-\xi_\gamma)\frac{p^\mu p^\nu}{p^2}\right],\\
\langle T~\phi^*_i(p) \phi_j(-p)\rangle &=&\Delta_{ij}(p)=\frac{i}{p^2}\delta_{ij}.
\end{eqnarray}
\noindent As we will restrict to calculations of processes with only external pion legs, at one loop order and at 
most with the exchange of one graviton, we will not need the ghost propagators.

We will consider two scattering processes: the first is ($\pi_a^+ +\pi_b^+\rightarrow \pi_a^+ +\pi_b^+$), that contains the exchange of photons only in the t-channel and the second is ($\pi_a^+ +\pi_a^-\rightarrow \pi_b^+ +\pi_b^-$) which contains photon propagators only in the s-channel.  These choices will allow a comparison with the results obtained in~\cite{Anber:2010uj}.  
 
The Lagrangian for the higher order monomials $\mathcal{L}_{HO} $ induced by quantum corrections consists of all possible monomials in the fields and their derivatives respecting the original symmetries of the model. The terms that are important in our analysis, i.e., which will be necessary to absorb the divergences appearing in the two reactions studied are
\begin{eqnarray}\label{eqho}
\mathcal{L}_{HO} &=& ie_1A^\nu\left[\partial^\mu\phi_j\partial_{\mu}\partial_{\nu}\phi^*_j-\partial^\mu\partial_{\nu}\phi_j\partial_{\mu}\phi^*_j\right]
+\lambda_1 \partial^\mu(\phi_a\phi^*_a)\partial_{\mu}(\phi_b\phi^*_b)\nonumber\\
&&+\lambda_2 (\phi_a \partial^\mu \phi^*_a-\partial^\mu\phi_a \phi^*_a)(\phi_b \partial_\mu \phi^*_b-\partial_\mu\phi_b \phi^*_b)+(\cdots),
\end{eqnarray}
\noindent where $(\cdots)$ stands for omitted higher order monomials, which are not important to our analysis.

\section{Scattering amplitudes and the running of the coupling constants} \label{sec2}

\begin{figure}[h]
\includegraphics[width=14cm]{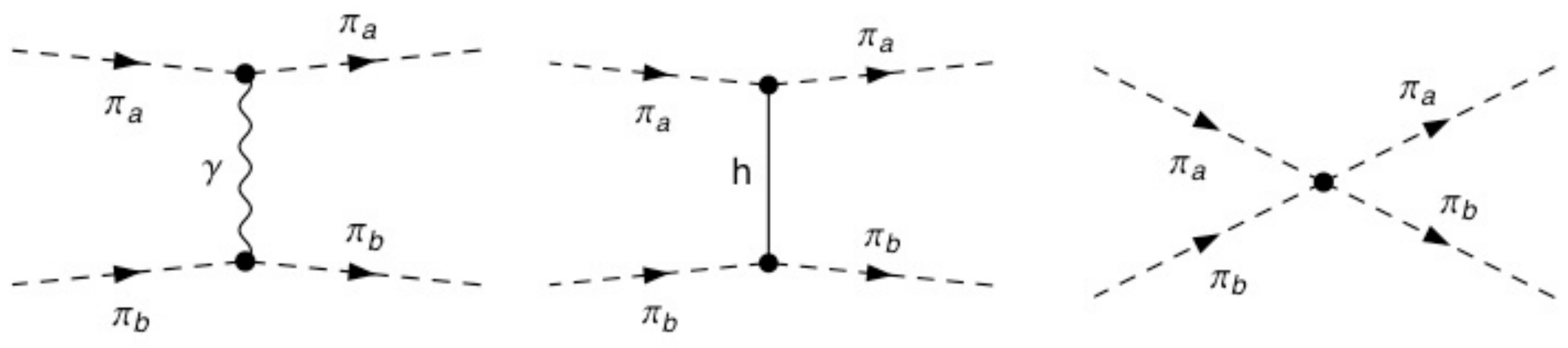} 
\caption{Feynman diagrams for the pions scattering amplitude ($\pi_a^+ +\pi_b^+\rightarrow \pi_a^+ +\pi_b^+$) at tree level. Dashed, wavy and continuous lines represent the scalar, photon and graviton propagators, respectively.}
\label{fig0}
\end{figure}

First, let us consider the process $\pi_a^+ +\pi_b^+\rightarrow \pi_a^+ +\pi_b^+$. The tree level amplitude, Fig. \ref{fig0}, including the contributions of the counterterms and of the higher order monomials, is given by
\begin{eqnarray}\label{sec2eq01}
\mathcal{M}_{1~tree}&=&-(\lambda+\delta_\lambda) +(e^2+2e\delta_e)\frac{S-U}{T}+\frac{\kappa^2+2\kappa\delta_\kappa}{4}\frac{US}{T}\nonumber\\ 
&& + (e~e_1+\lambda_2+e\delta_{e_1}+\delta_{\lambda_2})(S-U)+(\lambda_1+\delta_{\lambda_1})~T,
\end{eqnarray}

\noindent where $\delta_e$, $\delta_\lambda$, $\delta_\kappa$, $\delta_{e_1}$, $\delta_{\lambda_1}$ and $\delta_{\lambda_2}$ are the counterterms and the Mandelstam variables $S=(\text{p}_1+\text{p}_2)^2$, $T=(\text{p}_1-\text{p}_3)^2$ and $U=(\text{p}_1-\text{p}_4)^2$ are functions of the external incoming ($\text{p}_1$, $\text{p}_2$) and outgoing ($\text{p}_3$, $\text{p}_4$) momenta respectively, satisfying the relation $S+T+U=0$ for the massless pions on shell.
  
\begin{figure}[h]
\includegraphics[width=16cm]{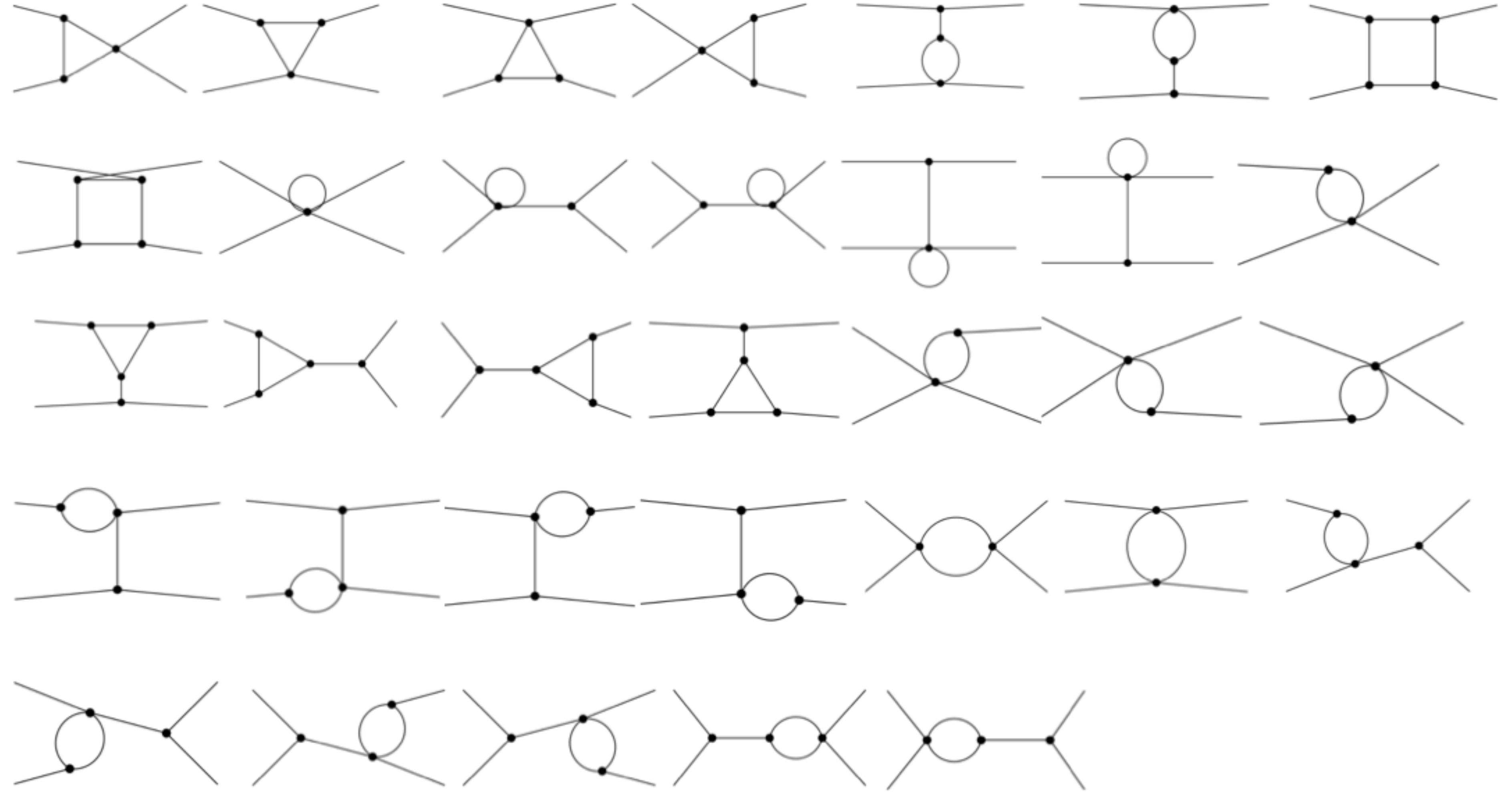} 
\caption{Topologies from  ($\pi_a^+ +\pi_b^+\rightarrow \pi_a^+ +\pi_b^+$) scattering amplitude at one-loop order. After to consider the vertices, we have more than 100 Feynman diagrams to this reaction up to ${\mathcal{O}}(\kappa^2)$.}\label{fig1}
\end{figure}

The one-loop correction to this process, Fig. \ref{fig1}, is given by
\begin{eqnarray}\label{sec2eq02}
\mathcal{M}_{1~lc}&=&-\frac{e^4}{2 \pi^2\epsilon}\frac{S-U}{T}+\frac{3 \lambda^2+3e^4-\lambda e^2}{4 \pi^2 \epsilon }-\frac{e^2\kappa^2}{8\pi^2\epsilon}\frac{US}{T}\nonumber\\
&&-\frac{\kappa^2(\lambda-2e^2)}{32\pi^2\epsilon}T-\frac{13e^2\kappa^2}{96\pi^2\epsilon}(S-U)+\text{finite~terms}  
\end{eqnarray}

\noindent where $\epsilon=4-D$, with $D$ being the dimension of space-time, in DR. To evaluate the amplitudes we used a set of Mathematica$^{\copyright}$ packages~\cite{feynarts,feyncalc,formcalc}.

The total amplitude $\mathcal{M}_1$ is given by the sum of the tree and the one-loop contributions, Eqs. (\ref{sec2eq01}) and (\ref{sec2eq02}). By eliminating one of the Mandelstam variables (say U) in terms of the others two and  imposing that the coefficient of any monomial (in S and T) is finite we get the renormalized amplitude, for the pions on the mass shell, for any values of the kinematical variables S and T.   
In the minimal subtraction scheme~\cite{minsub} the counterterms are given by
\begin{eqnarray}\label{ct01}
\delta_e&=&\frac{e^3}{4\pi^2\epsilon},\nonumber\\
\delta_\lambda&=&\frac{3\lambda^2+3e^4-\lambda e^2}{4\pi^2\epsilon},\nonumber\\
\delta_\kappa&=&\frac{\kappa e^2}{4\pi^2\epsilon},\\
\delta_{\lambda_1}&=&\frac{\kappa^2(\lambda-2e^2)}{32\pi^2\epsilon},\nonumber\\
e\delta_{e_1}+\delta_{\lambda_2}&=&\frac{13e^2\kappa^2}{96\pi^2\epsilon}.\nonumber
\end{eqnarray}

The renormalized amplitude results in:
\begin{eqnarray}\label{sec2eq012}
\mathcal{M}_{1}=&&-\lambda +e^2\frac{S-U}{T}+\frac{\kappa^2}{4}\frac{US}{T}
+ (e~e_1+\lambda_2)(S-U)\nonumber\\
&& +\lambda_1~T +\text{finite~terms}.
\end{eqnarray}

From the above expressions for the counterterms, we can obtain the corresponding beta functions of the dimensionless couplings $e$ and $\lambda$. Since $e_0=\mu^{\epsilon}Z_e e=\mu^{\epsilon}(e+\delta_e)$ and $\lambda_0=\mu^{\epsilon}Z_\lambda\lambda=\mu^{\epsilon}(\lambda+\delta_\lambda)$, where $\mu$ is the mass scale introduced in the DR procedure, their beta functions up to one-loop order are given by 
\begin{eqnarray}
\beta(e)&=&\frac{e^3}{4 \pi^2} \label{betae};\\
\beta(\lambda,e)&=&\frac{3 \lambda^2+3e^4-\lambda e^2}{4 \pi^2}\label{betal}.
\end{eqnarray}

What can be learned from this scattering amplitude is that quantum gravitational corrections only renormalize higher order operators, such as those in $\mathcal{L}_{HO}$, Eq.(\ref{eqho}). In addition, the electric charge and the the coupling constant $\lambda$ do not receive any gravitational correction and their beta functions are the same as in the absence of gravitation. Some results obtained by means of the method of effective action~\cite{Toms:2008dq,Pietrykowski:2012nc}, or scattering amplitudes~\cite{Lehum:2013oja}, suggest that this situation might change if we consider massive particles, or the presence of a positive cosmological constant.   
  
Instead of ($\pi_a^+ +\pi_b^+\rightarrow \pi_a^+ +\pi_b^+$), we could have used the ($\pi_a^+ +\pi_a^-\rightarrow \pi_b^+ +\pi_b^-$) process to obtain the same conclusions. By an explicit calculation, the tree level and the one-loop contributions to the scattering amplitude ($\pi_a^+ +\pi_a^-\rightarrow \pi_b^+ +\pi_b^-$) can be cast as
\begin{eqnarray}\label{sec2eq02a}
\mathcal{M}_{2}&=&-(\lambda+\delta_\lambda) +(e^2+2e\delta_e)\frac{U-T}{S}+\frac{\kappa^2+2\kappa\delta_\kappa}{4}\frac{UT}{S}\nonumber\\
&&+ (ee_1+\lambda_2+e\delta_{e_1}+\delta_{\lambda_2}) (U-T)+(\lambda_1+\delta_{\lambda_1})~S\nonumber\\
&& -\frac{e^4}{2 \pi^2\epsilon}\frac{U-T}{S}+\frac{3 \lambda^2+3e^4-\lambda e^2}{4 \pi^2 \epsilon }
-\frac{e^2\kappa^2}{8\pi^2\epsilon}\frac{UT}{S}\nonumber\\
&&-\frac{\kappa^2(\lambda-2e^2)}{32\pi^2\epsilon}S -\frac{13e^2\kappa^2}{96\pi^2\epsilon}(U-T)+\text{finite~terms}.
\end{eqnarray}
\noindent From this expression, we obtain the same equations (\ref{ct01}) for the counterterms, leading to the same conclusions about the running of the coupling constants. This is expected, because this amplitude can be gotten from the first one by the crossing symmetry: $\text{p}_1+\text{p}_2\rightarrow \text{p}_1-\text{p}_3$, $\text{p}_1-\text{p}_3\rightarrow \text{p}_1-\text{p}_4$ and $\text{p}_1-\text{p}_4\rightarrow \text{p}_1+\text{p}_2$, which implies:  $S \rightarrow T$, $T \rightarrow U$ and $U \rightarrow S$.

\section{Final Remarks}\label{summary}
 
In summary, we have studied the quantum gravitational corrections to the renormalization of coupling constants, of the massless Scalar Quantum Electrodynamics, through the evaluation of the scattering matrix for charged pions in the context of effective field theory. To do this, we have evaluated the UV divergent parts of the one-loop corrections to the S-matrix, up to order $\mathcal{O}(\kappa^2)$, using dimensional regularization and the minimal subtraction scheme of renormalization. We have shown that quantum gravitational corrections do not contribute to the running of the electric charge and the quartic self-coupling of the scalar fields. As is well known, dimensional regularization automatically renormalizes quadratic divergences. Quadratic divergences appear in some other regularization schemes, as for example, UV cutoff. But, as well clarified in ~\cite{Anber:2010uj}, quadratic divergences, after renormalization,  leave no trace in the renormalized (physical) coupling constants and so, the same conclusions as in DR regularization, must be expected. These results, do not preclude the possibility of change of the behavior of the coupling constants through logarithm divergences, when another dimensionful parameter is present in the theory. This could be so, because logarithms of a dimensionless parameter could be constructed from the two dimensionful parameters. 
 
The question about whether gravitational corrections should alter the running of the gauge (and $\phi^4$ interaction) constants, in the presence of others dimensionful parameters in the Lagrangian, remains open. As found by the computation of the effective action, logarithmic divergences induced by gravitational corrections can contribute to the beta function of the gauge couplings, when a cosmological constant~\cite{Toms:2008dq} or massive particles~\cite{Pietrykowski:2012nc} are included. The evaluation of scattering processes, should help to elucidate this problem. Another relevant question, no less controversial, is on the running of the gravitational coupling constant $\kappa$, which has been studied in~\cite{Weinberg:1980gg,Vacca:2010mj,Anber:2011ut} in  the context of asymptotic safety. The study of scattering processes of gravitons should also make an important contribution toward the clarification of these controversies. 

\vspace{.5cm}

{\bf Acknowledgments.}
This work was partially supported by Funda\c{c}\~ao de Amparo \`a Pesquisa do Estado de S\~ao Paulo (FAPESP), Conselho Nacional de Desenvolvimento Cient\'{\i}fico e Tecnol\'{o}gico (CNPq) and Funda\c{c}\~{a}o de Apoio \`{a} Pesquisa do Rio Grande do Norte (FAPERN). The authors would like to thank M. Gomes for the useful comments.

\newpage 

\begin{thebibliography}{99}

\bibitem{'tHooft:1974bx} 
  G.~'t Hooft and M.~J.~G.~Veltman,
  Annales Poincare Phys.\ Theor.\ A {\bf 20}, 69 (1974).
  
\bibitem{PhysRevLett.32.245} 
  S.~Deser and P.~van Nieuwenhuizen,
  Phys. Rev. Lett. {\bf 32}, 245 (1974).

\bibitem{Deser:1974cy} 
  S.~Deser and P.~van Nieuwenhuizen,
  Phys.\ Rev.\ D {\bf 10}, 411 (1974).

\bibitem{Donoghue:1994dn} 
  J.~F.~Donoghue,
  Phys.\ Rev.\ D {\bf 50}, 3874 (1994).
 
\bibitem{Robinson:2005fj} 
  S.~P.~Robinson and F.~Wilczek,
  Phys.\ Rev.\ Lett.\  {\bf 96}, 231601 (2006).

\bibitem{Pietrykowski:2006xy} 
  A.~R.~Pietrykowski,
  Phys.\ Rev.\ Lett.\  {\bf 98}, 061801 (2007).
  
\bibitem{Felipe:2012vq} 
J.~C.~C.~Felipe, L.~C.~T.~Brito, M.~Sampaio and M.~C.~Nemes,
  Phys.\ Lett.\ B {\bf 700}, 86 (2011);
  J.~C.~C.~Felipe, L.~A.~Cabral, L.~C.~T.~Brito, M.~Sampaio and M.~C.~Nemes,
  Mod.\ Phys.\ Lett.\ A {\bf 28}, 1350078 (2013).
  
\bibitem{Toms:2010vy} 
  D.~J.~Toms,
  Nature {\bf 468}, 56 (2010).

\bibitem{Toms:2011zza} 
  D.~J.~Toms,
  Phys.\ Rev.\ D {\bf 84}, 084016 (2011).

\bibitem{Nielsen:2012fm} 
  N.~K.~Nielsen,
  Annals Phys.\  {\bf 327}, 861 (2012).
  
\bibitem{Toms:2008dq} 
  D.~J.~Toms,
  Phys.\ Rev.\ Lett.\  {\bf 101}, 131301 (2008).

\bibitem{Rodigast:2009zj} 
  A.~Rodigast and T.~Schuster,
  Phys.\ Rev.\ Lett.\  {\bf 104}, 081301 (2010).

\bibitem{Anber:2010uj} 
  M.~M.~Anber, J.~F.~Donoghue and M.~El-Houssieny,
  Phys.\ Rev.\ D {\bf 83}, 124003 (2011).
  
 \bibitem{Donoghue:2012zc} 
  J.~F.~Donoghue,
  AIP Conf.\ Proc.\  {\bf 1483}, 73 (2012).

\bibitem{minsub} 
G.~'t Hooft,
  Nucl.\ Phys.\ B {\bf 61}, 455 (1973); 
  S.~Weinberg,
  Phys.\ Rev.\ D {\bf 8}, 3497 (1973).

\bibitem{Choi:1994ax} 
  S.~Y.~Choi, J.~S.~Shim and H.~S.~Song,
  Phys.\ Rev.\ D {\bf 51}, 2751 (1995).

\bibitem{feynarts} 
  Thomas Hahn,
  Comp. Phys. Comm. {\bf 140}, 418 (2001).
  
\bibitem{feyncalc} 
  R. Mertig and M. Bahm and A. Denne,
  Comp. Phys. Comm. {\bf 64}, 345 (1991).

\bibitem{formcalc} 
  T.~Hahn and M.~Perez-Victoria,
  Comput.\ Phys.\ Commun.\  {\bf 118}, 153 (1999).
 
 \bibitem{Pietrykowski:2012nc} 
  A.~R.~Pietrykowski,
  Phys.\ Rev.\ D {\bf 87}, 024026 (2013).

 \bibitem{Lehum:2013oja} 
  A.~C.~Lehum,
  Phys.\ Rev.\ D {\bf 88}, 104030 (2013).
  
\bibitem{Weinberg:1980gg} 
  S. Weinberg, In General Relativity: An Einstein Centenary Survey, ed. S.W. Hawking and W. Israel, pp.790-831; Cambridge University Press (1979).
  
\bibitem{Vacca:2010mj} 
  G.~P.~Vacca and O.~Zanusso,
  Phys.\ Rev.\ Lett.\  {\bf 105}, 231601 (2010).
  
\bibitem{Anber:2011ut} 
  M.~M.~Anber and J.~F.~Donoghue,
  Phys.\ Rev.\ D {\bf 85}, 104016 (2012).
  
\end{thebibliography}
\end{document}